\documentclass[aps,reprint,superscriptaddress,longbibliography,prb]{revtex4-1}

\usepackage{graphicx, amsmath, verbatim, dsfont, amsfonts, color}
\usepackage{braket}

\begin{document}

\title{Electrostatic control of quantum Hall ferromagnetic transition, a step toward reconfigurable network of helical channels}

\author{Aleksandr~Kazakov}
\affiliation{Department of Physics and Astronomy, Purdue University, West Lafayette, IN 47907 USA}
\author{George~Simion}
\affiliation{Department of Physics and Astronomy, Purdue University, West Lafayette, IN 47907 USA}
\author{Yuli~Lyanda-Geller}
\affiliation{Department of Physics and Astronomy, Purdue University, West Lafayette, IN 47907 USA}
\affiliation{Birck Nanotechnology Center, Purdue University, West Lafayette, IN 47907 USA}
\author{Valery~Kolkovsky}
\affiliation{Institute of Physics, Polish Academy of Sciences, Al. Lotnikow 32/46, 02-668 Warsaw, Poland}
\author{Zbigniew~Adamus}
\affiliation{Institute of Physics, Polish Academy of Sciences, Al. Lotnikow 32/46, 02-668 Warsaw, Poland}
\author{Grzegorz~Karczewski}
\affiliation{Institute of Physics, Polish Academy of Sciences, Al. Lotnikow 32/46, 02-668 Warsaw, Poland}
\author{Tomasz~Wojtowicz}
\affiliation{Institute of Physics, Polish Academy of Sciences, Al. Lotnikow 32/46, 02-668 Warsaw, Poland}
\author{Leonid~P.~Rokhinson}
\email{leonid@purdue.edu}
\affiliation{Department of Physics and Astronomy, Purdue University, West Lafayette, IN 47907 USA}
\affiliation{Birck Nanotechnology Center, Purdue University, West Lafayette, IN 47907 USA}
\affiliation{Department of Electrical and Computer Engineering, Purdue University, West Lafayette, IN 47907 USA}


\begin{abstract}
Ferromagnetic transitions between quantum Hall states with different polarization at a fixed filling factor can be studied by varying the ratio of cyclotron and Zeeman energies in tilted magnetic field experiments. However, an ability to locally control such transitions at a fixed magnetic field would open a range of attractive applications, e.g. formation of a reconfigurable network of one-dimensional helical domain walls in a two-dimensional plane. Coupled to a superconductor, such domain walls can support non-Abelian excitation. In this article we report development of heterostructures where quantum Hall ferromagnetic (QHFm) transition can be controlled locally by electrostatic gating. A high mobility two-dimensional electron gas is formed in CdTe quantum wells with engineered placement of paramagnetic Mn impurities. Gate-induced electrostatic field shifts electron wavefunction in the growth direction and changes overlap between electrons in the quantum well and d-shell electrons on Mn, thus controlling the s-d exchange interaction and the field of the QHFm transition. The demonstrated shift of the QHFm transition at a filling factor $\nu=2$ is large enough to allow full control of spin polarization at a fixed magnetic field.
\end{abstract}

\maketitle

\section{Introduction}

One of the key ingredient in the realization of topological superconductivity\cite{Kitaev2001} is to remove fermion doubling. The doubling is naturally absent in fully spin polarized systems, yet ferromagnetic spin arrangement is not compatible with a conventional s-wave superconductivity. It has been realized that spin doubling can be removed in spin-full systems if spin is locked to the carrier momentum\cite{fu08,Sau2010,Lutchyn2010a,Oreg2010,Alicea2010}. While signatures of Majorana fermions have been reported in hybrid semiconductor/superconductor nanowires \cite{Rokhinson2012a,Mourik2012,Churchill2013a}, removal of fermion doubling has been observed in electron transport only in the cleanest nanowires fabricated by cleaved edge overgrowth technique\cite{Quay2010}. 

An elegant proposal to circumvent fermion doubling is to couple two two-dimensional electron gases (2DEGs) with different sign of Land\'e $g$-factor and subject then to a quantized magnetic field\cite{Clarke2012,Stern2013}. In a quantum Hall effect (QHE) regime two oppositely polarized counter-propagating edge channels at the boundary of two 2DEGs form a helical domain wall ({\it h}-DW), similar to helical channels at the edges of two-dimensional topological isolators \cite{Hasan2010}. Coupled to an s-wave superconductor, {\it h}-DW should support Majorana fermions in the integer QHE regime and parafermions in the fractional QHE regime\cite{Clarke2012}.

\begin{figure}[t]
\centering\includegraphics[width=0.95\columnwidth]{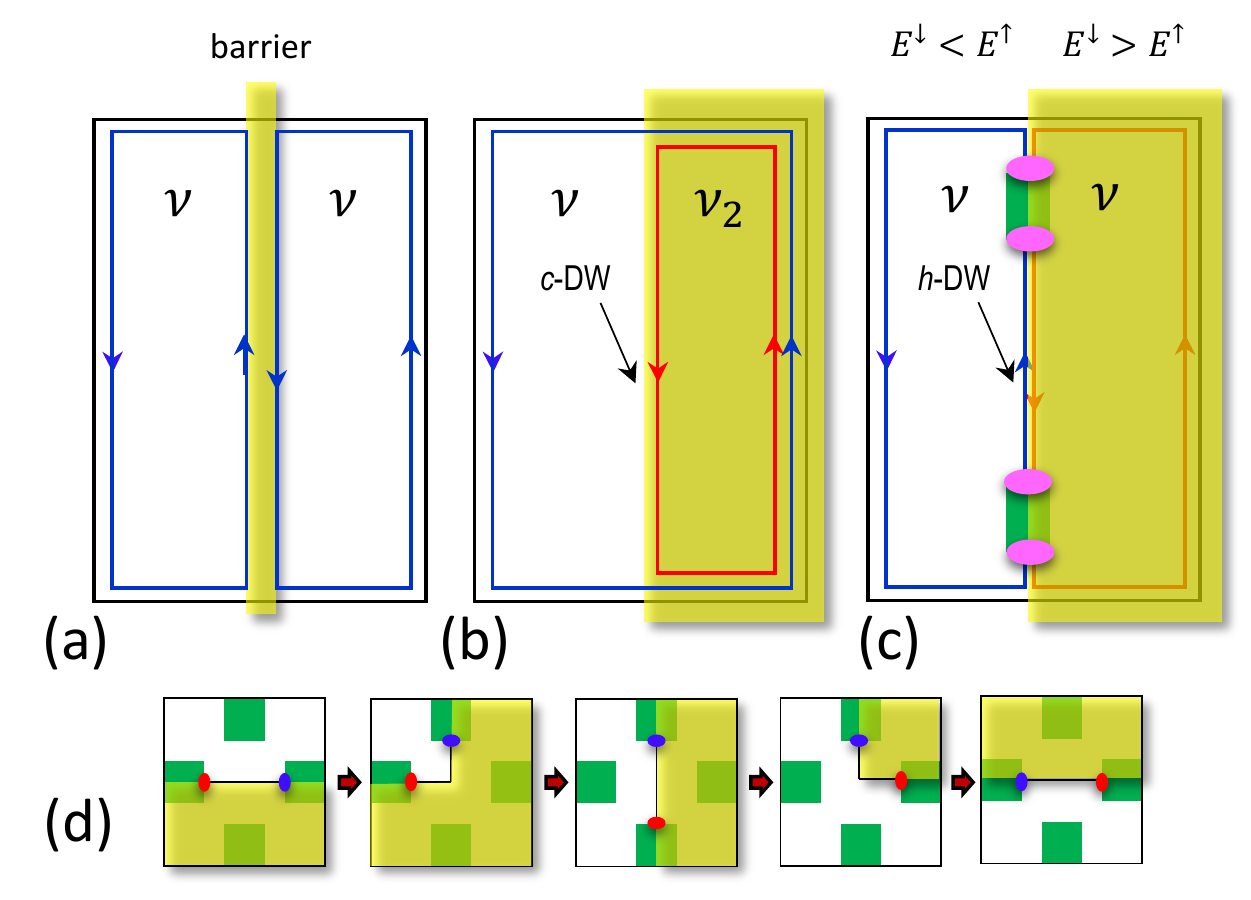}
\vspace{0in}
\caption{(a) In a QHE regime a potential barrier creates counter-propagating edge channels with the same polarizations, while (b) a filling factor gradient $\nu_2>\nu$ results in a formation of a chiral domain wall, \textit{c}-DW. (c) A local change of the topmost Landau level polarization results in the formation of a helical domain wall, \textit{h}-DW, where counter-propagating edge channels have opposite polarization. Coupled to a superconducting contact (green), these {\it h}-DW should support non-Abelian excitations (magenta dots). (d) Schematic of a reconfigurable {\it h}-DW in a multi-gate device.}
\label{f:helic}
\end{figure}

While bringing two different electron gases into a close proximity is an experimentally challenging proposition, we propose to use electrostatically controlled quantum Hall ferromagnetic (QHFm) transitions to form helical domain walls, see schematic in Fig.~\ref{f:helic}. In a QHE regime kinetic energy of electrons in a 2DEG is quantized into Landau levels (LL), which are further split due to the presence of spin. Polarization of a 2DEG and, more importantly, of the top filled energy level, depends on the number of occupied energy levels $\nu=n/n_\phi$ (the filling factor is a ratio of electron $n$ and magnetic flux $n_\phi=eB/h$ densities), and changes as the system undergoes phase transitions between QHE states with different filling factors. If a 2D gas is separated into regions with different $\nu$'s by, e.g., electrostatic gating, chiral current-caring states are formed at the boundary. The actual order of spin-split energy levels is determined by an intricate balance between Zeeman, cyclotron and exchange energies. By shifting the balance it is possible to induce magnetic phase transitions between different QHE states with the same filling factor. QHFm transitions in integer and fractional QHE regimes have been studied extensively in the past\cite{eisenstein90,DePoortere2000,Smet2001,Jaroszynski2002,Gusev2003,Betthausen2014}. The QHFm transition field $B^*(B_{||})$ in those experiments was adjusted by in-plane (Zeeman) magnetic field $B_{||}$, which does not afford local control of polarization.

In this article we report development and characterization of heterostructures where $B^*$ is sensitive to electrostatic gating, $B^*(V_g)$, and, thus, can be controlled locally, an enabling step toward experimental realization of theoretical concepts \cite{Clarke2012,Stern2013}. In devices with multiple gates a possibility to reconfigure a network of {\it h}-DW opens a new class of systems where non-Abelian excitation can be created and manipulated.

\section{Electrostatic control of quantum Hall ferromagnet}

\begin{figure}[t]
\centering\includegraphics[width=0.95\columnwidth]{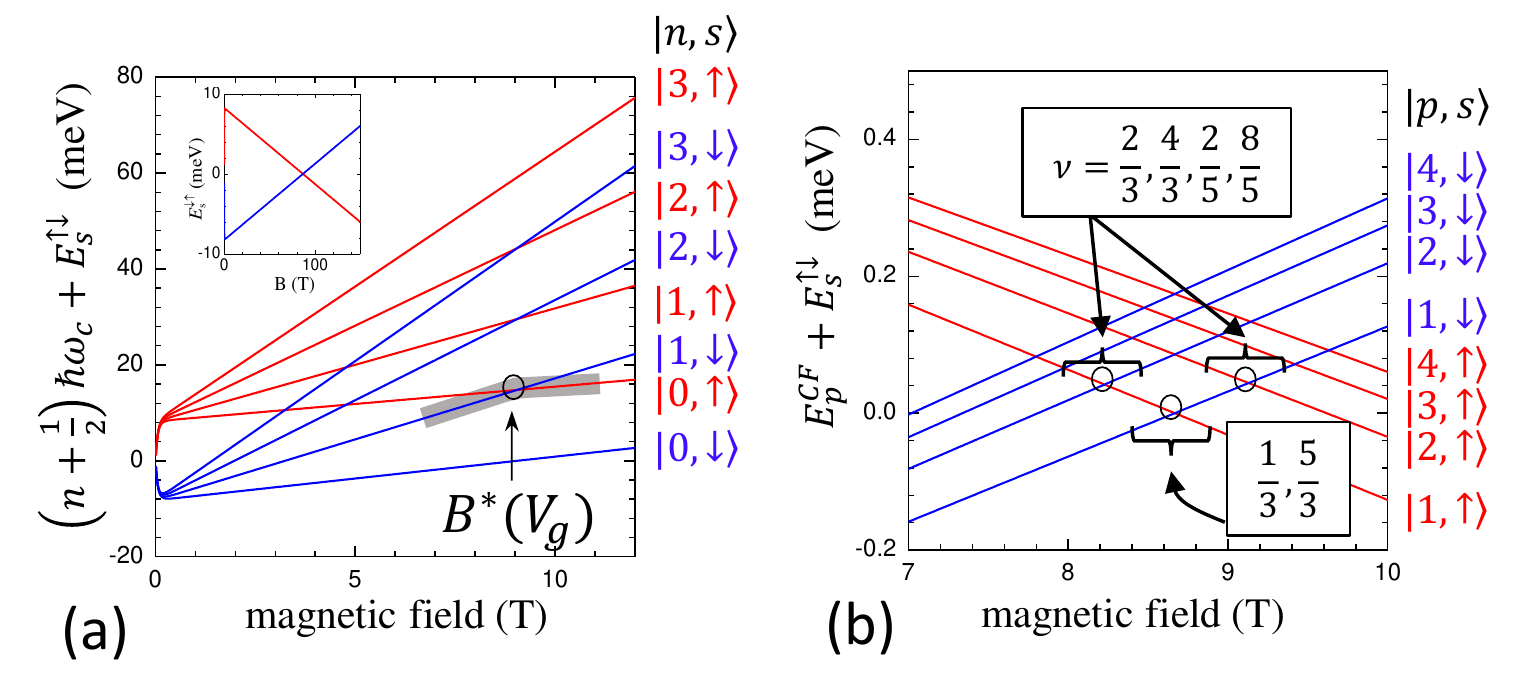}
\vspace{0in}
\caption{(a) Energy spectrum of Landau levels in a CdTe QW with 1.5\% of Mn calculated for $T=25$ mK. For the filling factor $\nu=2$ (gray shadow) electron gas undergoes ferromagnetic phase transition at $B^*(V_g)$. Field dependence of spin subbands (Eq.~\ref{eq:Es}) is plotted in the inset. (b) Spectrum for composite fermions $\Lambda$ levels for $x_{eff}=0.15\%$. QHFm transitions at $\nu=5/3$ and 4/3 have been experimentally observed\cite{Betthausen2014}.}
\label{f:effZ}
\end{figure}

\subsection{QHFm transition in dilute magnetic semiconductor}

Electrostatic control of QHFm transitions is realized in a dilute magnetic semiconductor CdTe:Mn with engineered placement of paramagnetic impurities. Substitutional Mn is a neutral impurity in CdTe and fractional QHE has been observed in high mobility CdTe:Mn two-dimensional electron gases with $\sim1\%$ of Mn\cite{Betthausen2014}. Exchange interaction between d-electrons on Mn (spin $S=5/2$) and s-electrons in the QW modifies energy spectrum of a 2DEG and  results in unusual spin splitting and level crossing at high magnetic fields\cite{Wojtowicz1999}. QHFm transition in both integer and fractional QHE regimes have been observed in tilted magnetic fields experiments in QWs with uniform Mn doping \cite{Jaroszynski2002,Betthausen2014}. In the presence of magnetic field $B$ spin-dependent energy in dilute magnetic semiconductors is \cite{Furdyna1988}:
\begin{equation}
\label{eq:Es}
E_s^{\uparrow\downarrow} = \pm\frac{1}{2}\left[g^*\mu_BB + x_{eff} E_{sd} S \mathfrak{B}_s\left(\frac{g^*\mu_BSB}{k_B(T+T_{AF})}\right)\right],
\end{equation}
where the first term is the Zeeman splitting and the second term is due to an s-d exchange. Here $g^*\approx-1.7$ in CdTe, $E_{sd}\approx 220$ meV \cite{Gaj1994,Kiselev1998}, $x_{eff}$ is an effective Mn concentration, and $T_{AF}$ is due to Mn-Mn antiferromagnetic interaction. At low fields spin splitting is dominated by a large positive exchange term, while at high fields and low temperatures the Brillouin function $\mathfrak{B}_s(B,T)\approx1$ and $B$-dependence is dominated by the negative Zeeman term. In Fig.~\ref{f:effZ} we plot spin splitting of energy levels (\ref{eq:Es}) and spectrum of Landau levels (LL) for electrons $(n+1/2)\hbar\omega_c+E_s^{\uparrow\downarrow}$ and composite fermions (CF) $E_p^{CF}+E_s^{\uparrow\downarrow}$, where energy gaps between CF levels\cite{Park1998} $E_{p+1}^{CF}-E_p^{CF}\approx \alpha_C E_c/(2p+1)\propto \sqrt{B}$. Here $\hbar$ is the reduced Plank's constant, $\omega_c$ is the cyclotron frequency, $E_c=e^2/\epsilon \ell$ is the charging energy, $\ell$ is the magnetic length, constant $\alpha_C\approx0.01-0.03$ depends on the confining potential \cite{Liu2014}, $n=0,1,2,...$ and $p=1,2,3...$. The field of spin subbands crossing $B^*$ for the same LL ($\ket{n,\uparrow}$ and $\ket{n,\downarrow}$) or neighboring LLs ($\ket{n,\uparrow}$ and $\ket{n\pm1,\downarrow}$) depends on the strength of the s-d exchange interaction $x_{eff} E_{sd}$. Thus, engineering heterostructures with gate-tunable s-d exchange will allow local control of spin polarization in both integer and fractional QHE regimes.

\subsection{Heterostructures with s-d exchange control}

The second term in (\ref{eq:Es}) is a mean-field approximation to the exchange Hamiltonian $J_{sd}\sum_{\vec{R}_i}\delta(\vec{r}-\vec{R}_i)\vec{S}_i\cdot \vec{\sigma}\propto \left[\int_{[Mn]}|\varphi(z)|dz\right] \langle\vec{S}\rangle$, where interaction of an electron at a position $\vec{r}$ with a large number of Mn ions at positions $\vec{R}_i$ is approximated as an overlap of the electron wavefunction $\varphi(z)$ with a uniform Mn background within $z\in[Mn]$ and an average magnetization $\langle\vec{S}\rangle=\langle S_z\rangle=S\mathfrak{B}_s(B,T)$. For quantum wells with homogeneous Mn distribution throughout the whole QW region an integral $\chi=\int_{[QW]}|\varphi(z)|dz$ has weak dependence on the shape of $\varphi(z)$ and level crossing field $B^*$ is almost independent of a gate voltage \cite{Jaroszynski2002}.

\begin{figure}[t]
\centering\includegraphics[width=0.95\columnwidth]{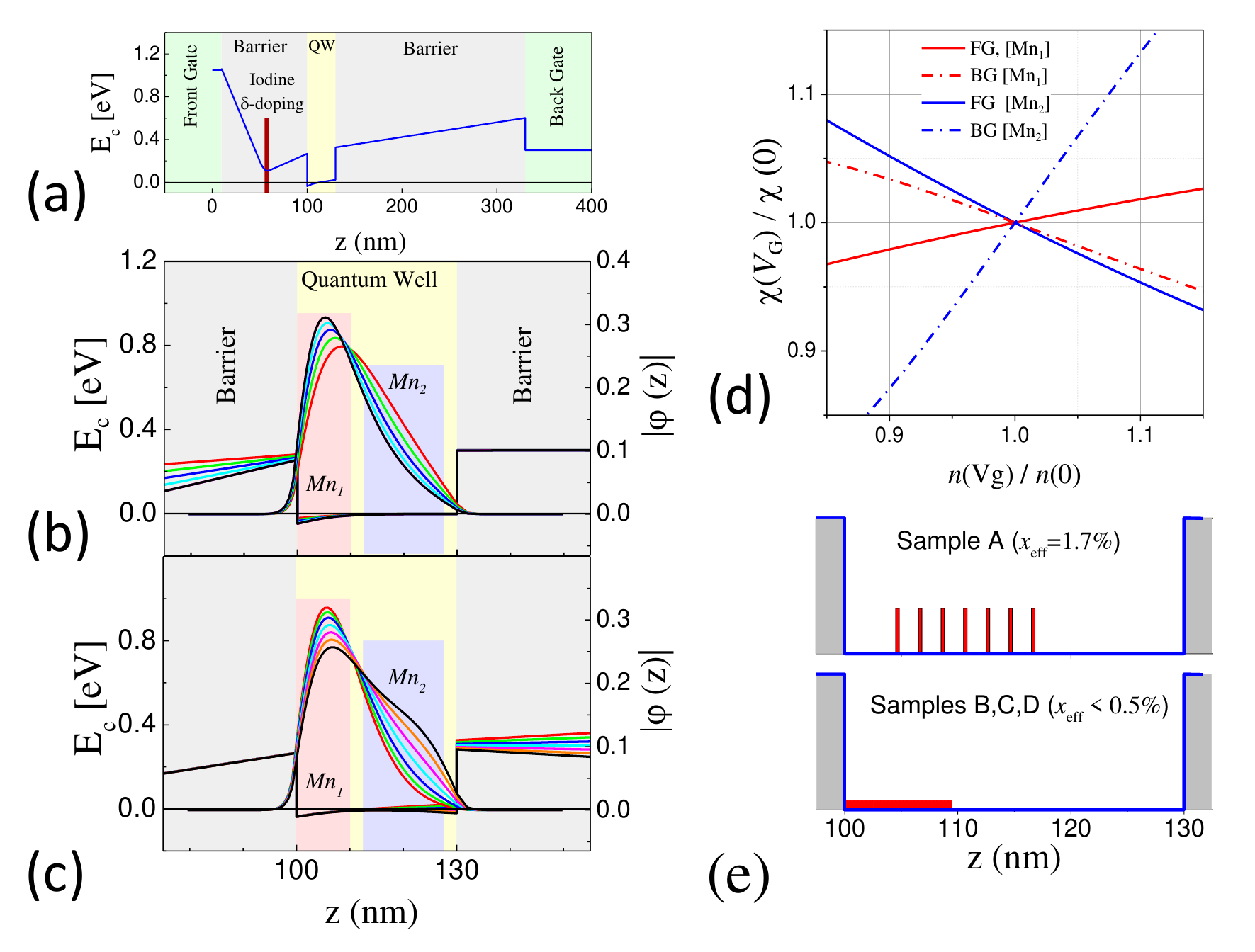}
\vspace{0in}
\caption{(a) Band diagram of a 30nm CdTe QW heterostructure device is modeled using next{\bf nano$^3$} package \cite{nextnano}. Electron wavefunction is calculated for different voltages on the top (b) and back (c) gates. In (d) an integral overlap $\chi(V_g)$ between Mn-doped regions $[Mn_1]$ and $[Mn_2]$, normalized to the value at zero gate voltage $\chi(0)$, is plotted as a function of the 2D gas density change for front (FG) and back (BG) gates. (e) Mn doping distribution (red regions) in different wafers.}
\label{bandeng}
\end{figure}

We now consider non-uniform distribution of Mn inside a QW, e.g. Mn is confined to regions $[Mn_1]$ or $[Mn_2]$ within the QW, see Fig.~\ref{bandeng}b,c. In these regions $\varphi(z)$ has strong dependence on the out-of-plane electric field and $\chi$ becomes gate dependent, $\chi=\chi(V_g)$. Application of positive (negative) voltage to the front gate shifts electron wavefunction closer to (away from) the surface, $d\chi/dV_{fg}>0$ for $[Mn_1]$ and $d\chi/dV_{fg}<0$ for $[Mn_2]$. Gate voltage also changes electron density $dn/dV_{fg}>0$, thus $d\chi/dn>0$ for $[Mn_1]$ and $d\chi/dn<0$ for $[Mn_2]$ for the front gate. Application of a back gate voltage results in a density change $dn/dV_{bg} > 0$ but electrical field shifts wavefunction in the opposite direction, thus $d\chi/dn<0$ for $[Mn_1]$ and $d\chi/dn>0$ for $[Mn_2]$ for the back gate. Described behaviour is summarized on a Fig.\ref{bandeng}d. For the formation of well defined {\it h}-DWs we want to control $B^*$ with a minimal change of $n$ in order to remain at the same filling factor $\nu$, or maximize $|d\chi/dn|$.

In order to demonstrate electrostatic control of QHFm transition several Cd$_{1-x}$Mn$_x$Te/ Cd$_{0.8}$Mg$_{0.2}$Te quantum well heterostructures were grown by molecular beam epitaxy (MBE), see \cite{Jaroszynski2002,Betthausen2014} for details. Iodine delta-doping layer is separated from the QW by a 30 nm Cd$_{0.8}$Mg$_{0.2}$Te spacer. Mn was introduced into the QW region either as a digital $\delta$-doping or as a continuous doping, see schematics in Fig.~\ref{bandeng}e. More than 35 wafers have been grown and characterized with different Mn placement and concentration, here we report data on 4 representative wafers with $x_{eff}=1.71\%$, 0.34\%, 0.20\% and 0.085\% (wafers A,B,C and D). Samples were patterned into $100 \mu$m-wide Hall bars. A semitransparent Ti front gate (10 nm thick) was thermally evaporated onto the central part of Hall bars. Ohmic contacts were produced by soldering freshly cut indium ingots similar to previous studies \cite{Jaroszynski2002, Betthausen2014}. Copper foil glued to the back of samples served as a back gate. Devices were illuminated with a red LED at 4 K, low temperature electron density and mobility were in the range of $3.2-3.5\cdot10^{11}$ cm$^{-2}$ and $2-3\cdot10^5$ cm$^2$/V$\cdot$s in different samples. Electron transport was measured in a dilution refrigerator using standard ac technique with 10 nA excitation.

\subsection{Smooth QHFm transition at $\nu=1$}

\begin{figure}[t]
\centering\includegraphics[width=0.95\columnwidth]{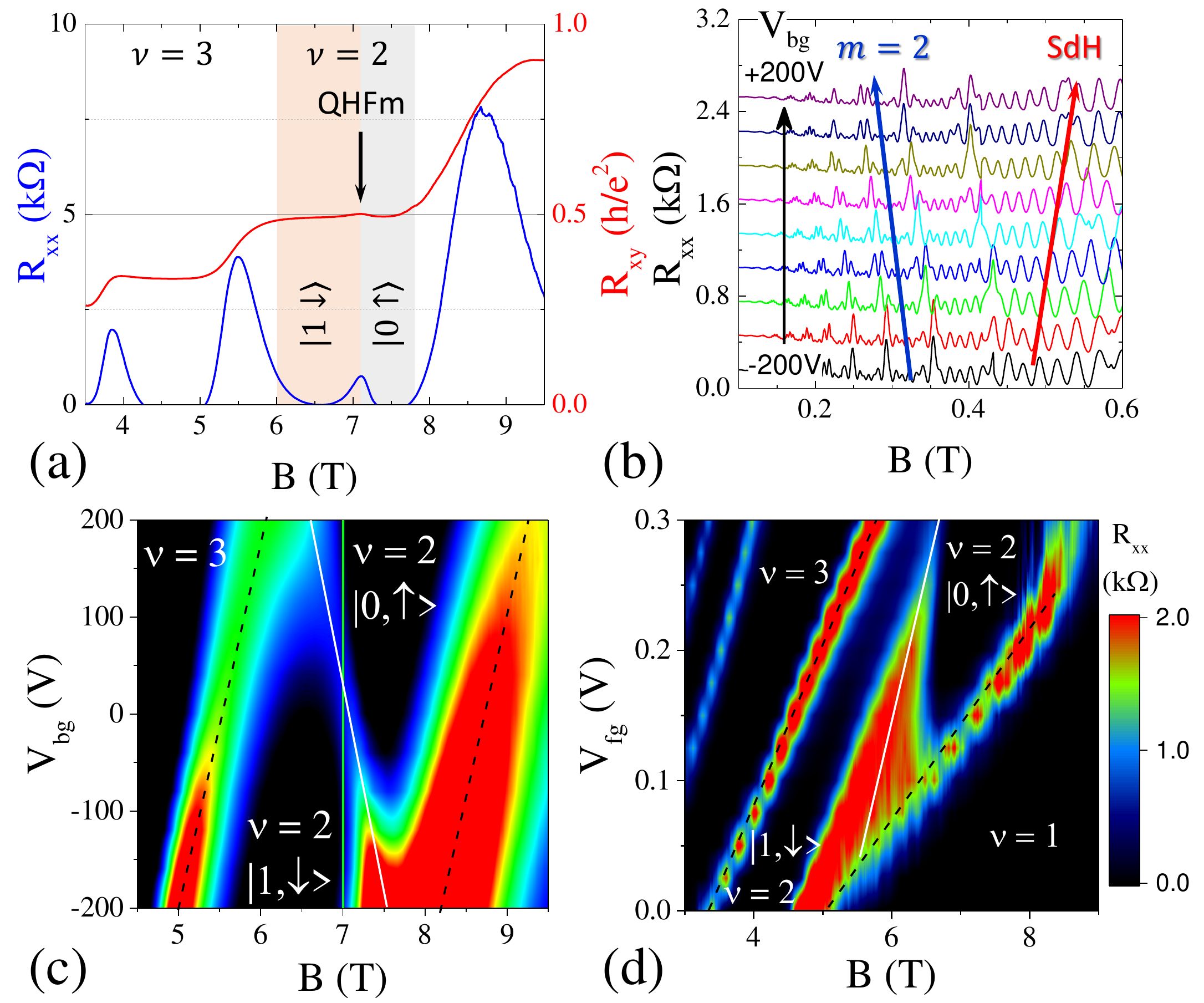}
\vspace{0in}
\caption{(a) Longitudinal ($R_{xx}$) and Hall ($R_{xy}$) magnetoresistances in wafer A measured at $T=400$ mK for $V_{fg}=V_{bg}=0$. A peak at $B=7$ T is a QHFm transition between $\ket{1\uparrow}$ and $\ket{0\downarrow}$ states. (b) magnetoresistance in wafer C measured at $T\approx30$ mK for various $V_{bg}$ from $-200$ V (bottom trace) to $+200$ V (top trace), the traces are offset proportional to $V_{bg}$. Blue arrow marks evolution of the $m=2$ node, red arrow marks evolution of SdH peaks. In (c) and (d) $R_{xx}$ in wafer A is plotted as a function of $V_{bg}$ or $V_{fg}$ at a fixed $V_{fg}=0$ or $V_{bg}=100$ V respectively. Position of the QHFm transition is highlighted by a white dotted line. For $B=7$ T polarization of the top LL can be switched between $\uparrow$ and $\downarrow$ by the gate. Both plots have the same color scale. Measurements are performed at $T=300$ mK.}
\label{f:gate}
\end{figure}

Spin levels crossing measured in optical experiments\cite{Wojtowicz1999} and QHFm transitions observed in the fractional QHE regime\cite{Betthausen2014} are well described by Eq.~\ref{eq:Es} and the values of $x_{eff}$ extracted from the beating of Shubnikov - de Haas (SdH) oscillations at low fields\cite{Teran2002}. Yet, we did not observe any re-entrant behavior at $\nu=1$. We conclude that the absence of a transport signature of the QHFm transition at $\nu=1$ is either due to a phase separation in the vicinity of the transition or strong e-e exchange interaction and anticrossing of levels with the same orbital wavefunction.

An ability to locally control exchange interaction for small $x_{eff}<0.01$ is crucial for the formation of {\it h}-DW in a fractional quantum Hall regime, a prerequisite for the creation of higher order non-Abelian excitations. The strength of the exchange interaction can be obtained from the beating in the SdH regime, where the $m$-th node is defined by the condition \cite{Teran2002} $(m+1/2)\hbar\omega_c=|E_s^{\uparrow}-E_s^{\downarrow}|$. Gate dependence of magnetoresistance in wafer C at low fields is shown in Fig.~\ref{f:gate}b. Nodes are shifted to lower fields as the voltage on the back gate increases, $d\chi/dV_{bg}<0$. At the same time SdH peaks shift to higher fields, $dn/dV_{bg}>0$, and $d\chi/dn<0$ as is expected for the [Mn$_1$] doping arrangement.

\subsection{Gate control of sharp QHFm transition at $\nu=2$}

Unlike $\ket{0\uparrow}\leftrightarrow\ket{0\downarrow}$ QHFm transition at $\nu=1$, the $\ket{0\uparrow}\leftrightarrow\ket{1\downarrow}$ transition at $\nu=2$ involves states from different Landau levels and e-e exchange is strongly suppressed. Also, at $\nu=2$ level crossing has much stronger $B$-dependence $\hbar\omega_c/B\approx1.6$ meV/T, as compared to $g\mu_B\approx0.057$ meV/T at $\nu=1$, which suppresses phase separation. As a result quantization is lifted in the vicinity of the QHFm transition and a prominent signature in magnetoresistance is observed\cite{Jaroszynski2002}.

Magnetoresistance in sample A is shown in Fig.~\ref{f:gate}a. A small peak at $B=7$ Tesla in the middle of the $\nu=2$ state is the QHFm phase transition between $\ket{1\downarrow}$ and $\ket{0\uparrow}$ states, polarization of the top filled energy level changes across the transition. In the color plots magnetoresistance is plotted as a function of voltage on the front and back gates (Fig.~\ref{f:gate}c,d), measurements are performed by sweeping magnetic field at constant gate voltages. Electron density increases with the increase of $V_{bg}$ and $V_{fg}$ and peaks between adjacent QHE states shift to higher $B$ in both plots. In contrast, the QHFm transition $B^*$ shifts in opposite directions as a function of $V_{fg}$ and $V_{bg}$, consistent with the modelling of the wavefunction-Mn$_1$ overlap $\chi(V_g)$ in Fig.~\ref{bandeng}d. Note that for $B=7$ Tesla polarization of the top level can be tuned between $\ket{1\downarrow}$ and $\ket{0\uparrow}$ states by electrostatic gating, thus realizing the theoretical concept of Fig.~\ref{f:effZ}a.

Gate control of the s-d exchange is summarized in Fig.~\ref{AC}a for several wafers. The absolute values of Mn concentration $x_{Mn}$ and s-d overlap $\chi$ cannot be measured independently with high accuracy, by a relative change of the exchange interaction can be obtained from the gate dependence of the experimentally measured $x_{eff}(V_g)/x_{eff}(0)=\chi(V_g)/\chi(0)$. Slopes $dx_{eff}(V_g)/dn(V_g)$ are in a good agreement with $d\chi(V_g)/dn(V_g)$ obtained from band simulations, Fig.~\ref{bandeng}d. Note that the efficiency of the s-d exchange control depends on the $|dx_{eff}/dn|$ slope: $dB^*/dn=d\chi/dn=dx_{eff}/dn$ for QHFm transitions in the integer QHE regime and $dB^*/dn\approx d\chi/dn$ in the fractional QHE regime for large fields.

\subsection{Spin-orbit-induced gap for $\nu=2$ QHFm transition}

\begin{figure}[t]
\centering\includegraphics[width=0.95\columnwidth]{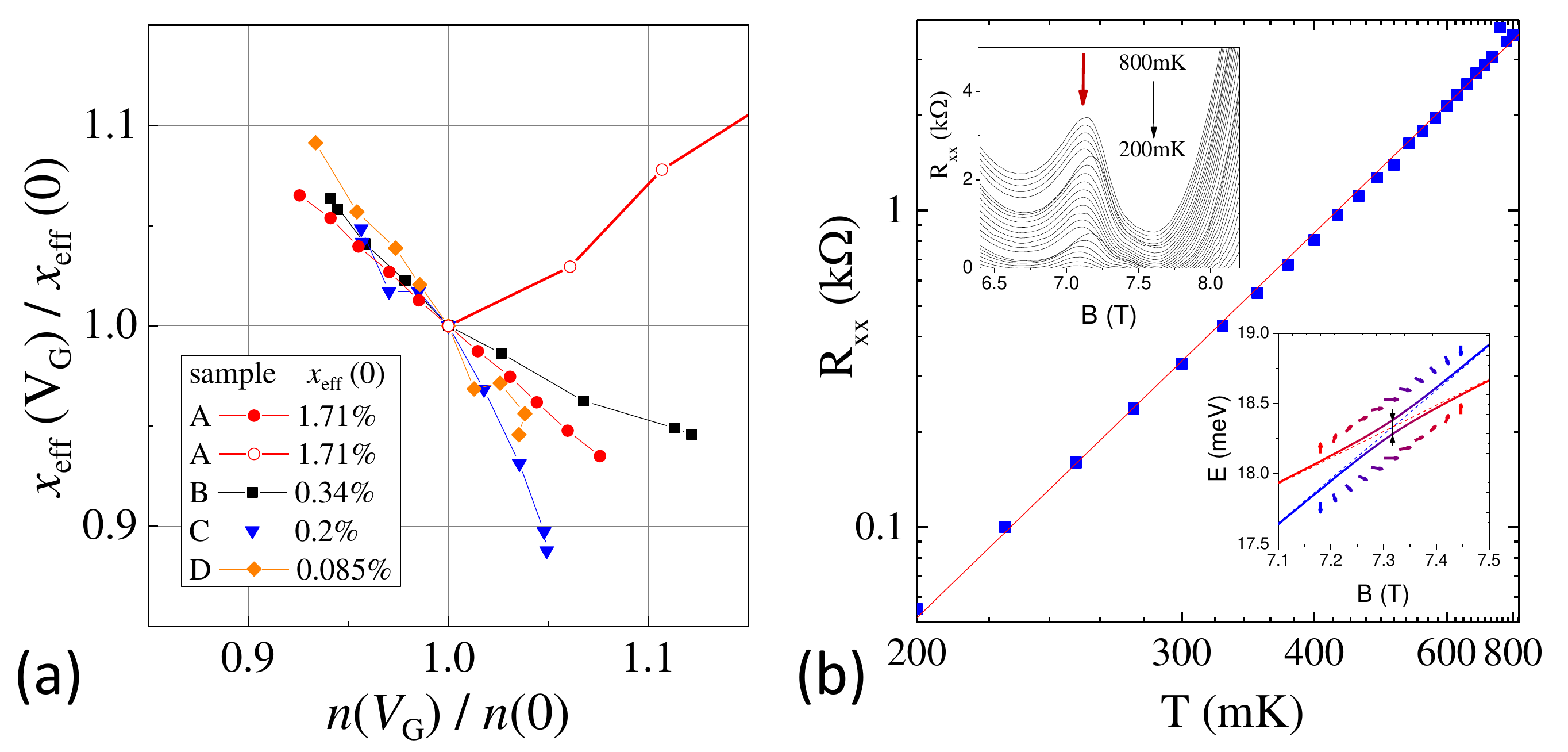}
\vspace{0in}
\caption{(a) Gate dependence of the measured effective Mn concentration, $\chi_{eff}(V_G)$ for wafers A-D for front (open symbols) and back (solid symbols) gates. Efficiency of s-d exchange control depend on the $|dx_{eff}/dn|$ slope: for QHFm transition in the integer QHE regime $(B^*(V_g)-B^*(0))\propto (x_{eff}(V_g)-x_{eff}(0))$. (b) Arrhenius plot of the $R_{xx}$ T-dependence at the QHFm transition, the activation energy is 0.096 meV. Top inset: temperature dependence of $R_{xx}$ near $\nu=2$. Bottom plot: anticrossing of $\ket{0,\uparrow}$ and $\ket{1,\downarrow}$ levels calculated using spin-orbit Hamiltonian, see text.}
\label{AC}
\end{figure}

The height of the peak at $B^*$ has exponential $T$-dependence and vanishes at low temperatures with an activation energy $T_0\approx 1K$, see Fig.~\ref{AC}b. We attribute this small gap to the level anticrossing due to the spin-orbit (SO) coupling between neighboring LLs. Energy spectrum in the presence of SO interactions is calculated by adding Dresselhaus $\gamma_D \boldsymbol{\kappa}\cdot\sigma$ and Rashba $\gamma_R {\boldsymbol{\mathcal E}}
\cdot \left(\boldsymbol{\sigma} \times \bf{k}\right)$ spin-orbit terms to the single-particle Hamiltonian of a 2D gas in magnetic field in the presence of s-d coupling (\ref{eq:Es}), square well confinement potential in $z$ direction, and electric field potential $e\phi(z)\approx e\mathcal E_z z $, see Appendix for  details. Here $\gamma_D$ and $\gamma_R$ are the Dresselhaus and Rashba constants, and $\boldsymbol{\kappa}$ is defined as $(\{\hat k_x,\hat k_y^2-\hat
k_z^2\}, \{\hat k_y,\hat k_z^2-\hat k_x^2\}, \{\hat k_z,\hat
k_x^2-\hat k_y^2\})$. The energy spectrum near $\ket{0\uparrow}$ and $\ket{1\downarrow}$ levels crossing is plotted in the insert in Fig.~\ref{AC}b. The value of the anticrossing gap is found to depend only on the Rashba spin-orbit coupling
\begin{equation}
\Delta_{SO}=\frac{2\sqrt{2} |\gamma_R \langle{\mathcal E}_z\rangle|}{ \ell}.
\label{SO}
\end{equation}
For an average electric field of $\langle{\mathcal E}_z\rangle=3.5\cdot 10^4~ {\rm{V/cm}} $, $B=7~ {\rm T}$ and $\gamma_R=6.9~ e{\rm{\AA}}^2$ the calculated gap $\Delta_{SO}=70$ $\mu$eV, in a good agreement with the experimentally measured activation gap of 96 $\mu$eV. We note that an ability to open a topologically trivial (spin-orbit) gap is required for the localization of non-Abelian excitations \cite{Clarke2012}.

\section{Conclusions}

In this paper we propose a new experimentally feasible platform to realize non-Abelian excitations. The platform is based on the ability to create ferromagnetic domains in a quantum Hall effect regime, where helical domain walls are formed at the domain boundaries. These domain walls, coupled to a superconductor with high critical field $B_c$, should support Majorana and higher order non-Abelian excitations. Topological protection of the QHE regime insures that only single channel with removed fermion doubling is formed, thus alleviating multi-channel complication encountered in nanowire-based devices.  As a proof-of-concept we developed CdTe quantum well heterostructures with engineered placement of paramagnetic Mn impurities and demonstrated local control of the QHFm transition at $\nu=2$ by electrostatic gating. Further research is needed to develop superconducting contacts to CdTe, a possible path is to overgrow CdTe with HgCdTe/HgTe epilayers where ohmic contacts with a high-$B_c$ superconductor Nb have been demonstrated\cite{Oostinga2013}.

\section{Acknowledgements}

Authors acknowledge support by the U.S. Department of Energy, Office of Basic Energy Sciences, Division of Materials Sciences and Engineering under Awards DE-SC0008630 (A.K and L.P.R.), by the Department of Defence Office of Naval research Award N000141410339 (A.K, T.W., G.S. and Y. L-G.), by the National Science Centre (Poland) grant DEC-2012/06/A/ST3/00247 (V.K., Z.A., G.K., and T.W.), and by the Foundation for Polish Science (T.W.).

\begin{widetext}
\appendix*

\section{Calculation of spin-orbit--induced anticrossing of LLs in the presence of s-d exchange}

A general single-particle Hamiltonian can be written as:
\begin{eqnarray}
\label{eq:H_gen} H_0=& \frac{1}{2m^*}\left(-i \hbar \boldsymbol
\nabla+\frac{e}{c}{\bf{A}} \right)^2+
\frac{1}{2}g\mu_{B}{\bf{B}}\cdot\boldsymbol{\sigma}-e\phi({\bf{r}})+V_b(z) \\ \nonumber
&-J\boldsymbol{\sigma}\cdot\sum_{i}{{\bf{S_i}}\delta\left({\bf{R}}_{i}-{\bf{r}}\right)}+\gamma_D
\boldsymbol{\kappa}\cdot\sigma +\gamma_R {\boldsymbol{\mathcal E}}
\cdot \left(\boldsymbol{\sigma} \times \bf{k}\right)~,
\end{eqnarray}
where $\boldsymbol{\sigma}$ is a vector containing Pauli matrices,
$\phi$ is an electric potential, and $V_b$ is a confinement potential in $z$ direction, the fifth term is s-d exchange with Mn impurities and the last two terms are Dresselhaus and Rashba spin-orbit coupling. $\boldsymbol{\kappa}$ is defined as $(\{\hat k_x,\hat k_y^2-\hat
k_z^2\}, \{\hat k_y,\hat k_z^2-\hat k_x^2\}, \{\hat k_z,\hat
k_x^2-\hat k_y^2\})$, where $\{A,B\}=(AB+BA)/2$ and
${\bf{k}}=-i\hbar\boldsymbol{\nabla}+e{\bf{A}}/c$. Magnetic
field ${\bf{B}}=(0,0,B)$ corresponds to a vector potential
${\bf{A}}=(0, B x, 0)$.

For exchange interaction we use a mean-field model described in the main text
$-J\rho_i \langle{\bf{S}}\rangle\cdot \boldsymbol{\sigma}=-J \rho_i
\mathcal B_S(g\mu_B S |B|/ k_B T) \boldsymbol{\sigma} \cdot
{\bf{b}}$ , where $\rho_i$ is the density of ions, $\mathcal B$
is the Brillouin function, $k_B$ is Boltzmann's constant, and ${\bf{b}} $ is a unit vector in the direction of
magnetic field. We consider high-field limit $\mathcal B_S(x)=1$. We also assume uniform Mn doping in the range $z_{min}<z<z_{max}$. Electric potential
is $\phi(x,z)\approx -\mathcal E_z z $, we consider $\mathcal
E_z>0$.

The Hamiltonian describing motion in $z$ direction is
\begin{equation}
H_z=-\frac{\hbar^2}{2m}\frac{\partial^2}{\partial z^2} +V_b(z)-e
|\mathcal E_z| z-J\theta(z-z_{min})\theta(z_{max}-z)\sigma_z~.
\end{equation}
Its eigenvalues for the lowest subband are:
\begin{eqnarray}
\label{eq:lambda_ud} \Lambda_s&=&e |\mathcal E_z| \left[-w
-\left(\frac{\hbar^2}{2m e |\mathcal
E_z|}\right)^{\frac{1}{3}}\rm{ai}_1\right]\\ \nonumber
&+&\frac{sJ}{{\rm{Ai}}'\left({\rm{ai}}_1\right)^2} \left\{
{\rm{Ai}}'[\Xi\left(z_{min}\right)]^2- {\rm{Ai}}'[\Xi(z_{max})]^2-
{\rm{Ai}} \left[\Xi\left(z_{min}\right)\right]^2 \Xi\left(z_{min}
\right) +
 {\rm{Ai}} \left[\Xi\left(z_{max}\right)\right]^2
\Xi\left(z_{max} \right) \right\}
\end{eqnarray}
where $\rm{Ai}$ is the Airy function, $\rm{ai_1}$ is its first zero,
$\Xi(z)=(w+z)(2 m e|\mathcal E_z|/\hbar^2)^{1/3}+{\rm{ai}}_1$, and
$s=1$ for spin up states and $s=-1$ for spin down.

In the presence of perpendicular magnetic field an effective Hamiltonian is
\begin{equation}
\label{eq:H02D} H_0= \left(
\begin{array}{cc}
\hbar \omega_C \left(a^{\dagger}a^{\dagger}+\frac{1}{2}\right) +
\frac{1}{2}\left(g\mu_B B+\delta \Lambda\right) &
i\frac{\gamma_D}{\sqrt{2}\ell^3}\left(\hat a^{\dagger} \hat a \hat
a^{\dagger} -a^3-2
\ell^2 k_z^2 \hat a^{\dagger}\right)+\sqrt{2}\frac{\gamma_R |\mathcal E_z|}{\ell} \hat a \\
-i\frac{\gamma_D}{\sqrt{2}\ell^3 }\left[\hat a \hat a^{\dagger} \hat
a -(\hat a^{\dagger})^3 -2 \ell^2 k_z^2 \hat
a\right]+\sqrt{2}\frac{\gamma_R |\mathcal E_z|}{\ell}\hat
a^{\dagger} & \hbar \omega_C
\left(a^{\dagger}a^{\dagger}+\frac{1}{2}\right) -
\frac{1}{2}\left(g\mu_B B+\delta \Lambda\right)
\end{array}
\right)~,
\end{equation}
where lowering and rising operators are defined as
$a^{\dagger}=(\hat k_y-i\hat k_x)/\sqrt{2}$, $a=(\hat k_y+i\hat
k_x)/\sqrt{2}$, $\ell=\sqrt{eB/\hbar}$ is the magnetic length, and
$\hbar \omega_C=\hbar eB/m$ is the cyclotron energy.

We treat spin-orbit couplings as perturbations and found that only Rashba
term has a non-zero matrix element between $\ket{0\uparrow}$ and
$\ket{1\downarrow}$ energy levels. In the vicinity of crossing the
energy spectrum is
\begin{equation}
E_{\pm}=\hbar\omega_C\pm \frac{1}{2}\sqrt {\left(\hbar \omega_C -
g\mu_B B-\delta\Lambda \right)^2+\frac{8 \gamma_R^2 \mathcal
E_z^2}{\ell^2}},
\end{equation}
and the anticrossing gap is given by Eq.~\ref{SO}.
\end{widetext}

\end{document}